\documentclass[conference]{IEEEtran}
\IEEEoverridecommandlockouts

\usepackage{cite}
\usepackage{amsthm,amsmath,amsfonts}
\usepackage{algorithmic}
\usepackage{graphicx}
\usepackage{textcomp}
\usepackage[table,xcdraw]{xcolor}

\usepackage[ruled,vlined,linesnumbered]{algorithm2e}
\usepackage{multirow}
\usepackage{adjustbox}
\usepackage{pifont}
\usepackage{threeparttable}
\usepackage{booktabs}
\usepackage{bbding}
\usepackage{tablefootnote}
\usepackage{hyperref}
\usepackage{tcolorbox}
\usepackage{enumitem}
\usepackage{enumerate}
\usepackage{subfig}
\usepackage{url}
\usepackage{soul}
\usepackage{balance}
\usepackage{makecell}

\usepackage{stfloats}
\usepackage{subcaption} %
\usepackage{listings}

\usepackage{xspace}
\usepackage{tikz}
\usepackage{color}
\usepackage{xparse}
\usepackage{realboxes}	%

\hypersetup{
colorlinks=true,
urlcolor=black,
linkcolor=black,
citecolor=black
}

\def\BibTeX{{\rm B\kern-.05em{\sc i\kern-.025em b}\kern-.08em
    T\kern-.1667em\lower.7ex\hbox{E}\kern-.125emX}}

\begin{document}

\newcommand{\tech}{CLAST}

\newcommand{\utgen}{UTgen}
\newcommand{\origin}{Origin}

\newcommand{\raggen}{RAGGen}
\newcommand{\telpa}{TELPA}

\newcommand{\noslicing}{``\textit{w/o purify}''}
\newcommand{\nopa}{``\textit{w/o post}''}
\newcommand{\noboth}{``\textit{w/o both}''}

\newcommand{\noslicingname}{\textit{w/o purify}}
\newcommand{\nopaname}{\textit{w/o post}}
\newcommand{\nobothname}{\textit{w/o both}}

\newcommand{\SlateBlue}[1]{\textcolor[RGB]{199,21,133}{#1}}
\newcommand{\purple}[1]{\textcolor[RGB]{138,43,226}{#1}}

\newcommand{\yl}[1]{\SlateBlue{[Lin: #1]}}

\newcommand{\jj}[1]{{\color{orange}{Junjie: #1}}}

\newcommand{\wang}[1]{{\color{magenta}{Dong: #1}}}

\newcommand{\yc}[1]{{\purple{#1}}}

\newcommand{\colorblue}{\color{black}} %
\newcommand{\ins}[1]{{\colorblue{#1}}}

\newcommand{\reply}[1]{{\colorblue{#1}}}

\newcommand{\inlinecode}[1]{\colorbox{gray!10}{\lstinline|#1|}}

\title{Clarifying Semantics of In-Context Examples for Unit Test Generation
}

\author{
 \IEEEauthorblockN{
    Chen Yang\IEEEauthorrefmark{4}, 
    Lin Yang\IEEEauthorrefmark{4}, 
    Ziqi Wang\IEEEauthorrefmark{4}, 
    Dong Wang\IEEEauthorrefmark{4},
    Jianyi Zhou\IEEEauthorrefmark{2}, 
    Junjie Chen\IEEEauthorrefmark{4}\IEEEauthorrefmark{1}
     \thanks{\IEEEauthorrefmark{1} corresponding author}
}
\IEEEauthorblockA{
    \IEEEauthorrefmark{4}College of Intelligence and Computing, Tianjin University, Tianjin, China\\
    \IEEEauthorrefmark{2}Huawei Cloud Computing Technologies Co., Ltd., Beijing, China
}
\IEEEauthorblockA{
    \{yangchenyc, linyang, wangziqi123, dong\_w, junjiechen\}@tju.edu.cn, zhoujianyi2@huawei.com
}
}

\maketitle

\begin{abstract}
Recent advances in large language models (LLMs) have enabled promising performance in unit test generation through in-context learning (ICL). However, the quality of in-context examples significantly influences the effectiveness of generated tests—poorly structured or semantically unclear test examples often lead to suboptimal outputs. 
In this paper, we propose \tech{}, a novel technique that systematically refines unit tests to improve their semantic clarity, thereby enhancing their utility as in-context examples. 
The approach decomposes complex tests into logically clearer ones and improves semantic clarity through a combination of program analysis and LLM-based rewriting. 
We evaluated \tech{} on four open-source and three industrial projects. The results demonstrate that \tech{} largely outperforms \utgen{}, the state-of-the-art refinement technique, in both preserving test effectiveness and enhancing semantic clarity.
Specifically, \tech{} fully retains the original effectiveness of unit tests, while \utgen{} reduces compilation success rate (CSR), pass rate (PR), test coverage (Cov), and \ins{mutation score (MS)} by an average of 12.90\%, 35.82\%, 4.65\%, and \ins{5.07\%}, respectively.
Over 85.33\% of participants in our user study preferred the semantic clarity of \tech{}-refined tests. 
Notably, incorporating \tech-refined tests as examples effectively improves ICL-based unit test generation approaches such as RAGGen and TELPA, resulting in an average increase of 25.97\% in CSR, 28.22\% in PR, and 45.99\% in Cov for generated tests, compared to incorporating \utgen{}-refined tests.
The insights from the follow-up user study not only reinforce \tech{}'s potential impact in software testing practice but also illuminate avenues for future research.

\end{abstract}


\vspace{-1mm}
\section{Introduction}
\label{sec:intro}

Automated unit test generation is vital for enhancing software quality by producing tests to verify individual components. 
While search-based approaches that apply heuristic optimization methods to explore the test space have been extensively studied over the years~\cite{pynguin,evosuite,randoop}, recent advances in large language models (LLMs) offer a new paradigm. 
By learning from vast code repositories, 
LLMs can infer semantic relationships between code and its corresponding tests, enabling the generation of more context-aware unit tests that address the limitations of traditional methods and enhance overall test effectiveness.

In-Context Learning (ICL) has emerged as a key technique for harnessing LLMs' inference capabilities in automated test generation. 
Methods such as Retrieval-Augmented Generation (RAG) and few-shot learning enable LLMs to adapt to specific tasks using in-context test examples, producing more syntactically valid tests with improved coverage while eliminating the need for resource-intensive fine-tuning.
Prior work (i.e., \raggen{}~\cite{ase_empirical} and \telpa{}~\cite{telpa}) has demonstrated this potential.
For instance, \raggen{} retrieves unit tests corresponding to the methods similar to the focal method (i.e., method under test) as examples. \telpa{} selects a set of unit tests generated by a search-based approach as counter-examples to guide LLMs in generating more diverse tests. 
However, the effectiveness of these ICL-based approaches critically depends on the semantic clarity of the provided test examples. 
\ins{
Specifically, semantic clarity refers to how clearly a unit test conveys its purpose and behavior, comprising two aspects: (i) logical clarity: whether the test targets a single, well-defined scenario, as mixing unrelated assertions often leads to complex logic that obscures interpretability; and (ii) textual clarity: whether identifiers and code comments accurately describe the behaviors of test components.
}
Unfortunately, current literature has revealed that both developer-written and tool-generated unit tests suffer from semantic clarity issues, such as ambiguous identifiers and insufficient comments~\cite{ase_empirical,utgen}.
These issues in existing tests create a ``noisy curriculum'' for LLMs, limiting their ability to learn clear and meaningful patterns. 
Even worse, when examples fail to clearly articulate testing intent, such as mixing assertions for distinct behaviors in a single test, LLMs may inherit these ambiguities, resulting in generated tests with low coverage or logical errors. 
Hence, enhancing the semantic clarity of in-context test examples becomes pivotal to unlocking the full potential of ICL for test generation.

To address this limitation, our work focuses on the crucial aspect, i.e., the \textbf{semantic clarity} of in-context examples.
Drawing on insights from existing studies~\cite{ase_empirical,utgen, dan} and our empirical observations, two common factors have been identified that hinder semantic clarity of unit tests:  
(i) complex logic arising from multiple test scenarios within a single test, and (ii) insufficient textual clarity such as ambiguous identifiers or missing essential comments.
Intuitively, eliminating these issues can improve semantic clarity of unit tests, making it easier for LLMs to comprehend and learn from test examples.

Recently, Deljouyi et al.~\cite{utgen} proposed \utgen{}, which invokes LLMs to refine the unit tests generated by search-based tools to enhance their textual clarity, making them more semantically expressive.
However, its effectiveness is unsatisfactory for two main reasons:
on one hand, the complex logic of unit tests poses difficulties for LLMs to comprehend test semantics for test refinement with enhanced semantic clarity;
on the other hand, LLMs' hallucinations could compromise the original unit tests' effectiveness (e.g., test coverage and syntactic correctness) after refinement.
Embedding such non-effectiveness-preserving tests as examples could even negatively affect the effectiveness of such ICL-based test generation techniques, e.g., line coverage achieved by \raggen{} on the Time project is decreased from 57.32\% to 22.94\% after refining test examples with \utgen{} (Section~\ref{sec:results_rq2}).
Thus, enhancing semantic clarity of unit tests to improve ICL-based unit test generation remains a challenging and non-trivial endeavor.

In this work, we propose a novel unit test refinement technique, called \textbf{\tech{}} (\textbf{CLA}rifying \textbf{S}emantics of unit \textbf{T}ests), to enhance the semantic clarity of unit tests from the two aforementioned factors.
To address the limitation of the complex test logic resulting from multiple scenarios (i.e., multiple assertions with different purposes) within a single unit test, \tech{} splits a complex unit test into a set of purified ones, each of which describes a single test scenario, by slicing the test code from the assertion perspective.
To address the limitation of insufficient textual clarity, \tech{} leverages both LLMs and program analysis to refine identifiers and generate essential comments for each purified unit test.
Compared to complex unit tests, LLMs can better learn the semantics of purified tests, enabling them to generate more precise identifiers and comments.
Particularly, with the assistance of program analysis, \tech{} avoids potential errors caused by LLMs' hallucinations for achieving effectiveness-preserving test refinement.
Specifically, \tech{} identifies the generated comments and refined identifiers from the contents produced by LLMs for clarity enhancement based on Abstract Syntax Tree (AST) and textual analysis, and then integrates those refined information into the original code of the purified test via AST node matching.
Using \tech{}, each unit test is refined into a set of purified tests that preserve the original effectiveness but have semantically expressive comments and identifiers.
Embedding these refined tests as in-context examples could enable LLMs to glean more valuable knowledge, enhancing the effectiveness of ICL-based unit test generation.

To evaluate the effectiveness of \tech{}, we conducted an extensive study using seven real-world Java projects, including four open-source projects and three industrial projects.
We first evaluated whether \tech{} can refine unit tests more effectively compared to the state-of-the-art test refinement technique (i.e., \utgen{})~\cite{utgen} in terms of the degree to preserving the effectiveness of original tests.
Our results show that \utgen{} largely damages test effectiveness after refinement, with \ins{12.90}\%, \ins{35.82}\%, \ins{4.65}\%, and \ins{5.07\%} decrements in terms of compilation success rate (CSR), pass rate (PR), line coverage (Cov), \ins{and mutation score (MS)} respectively, while \tech{} completely preserves the effectiveness of original unit tests.
A subsequent user study demonstrated that the unit tests refined by \tech{} exhibited superior semantic clarity compared to both the original tests and those refined by \utgen{}.
Over 85.33\% of participants favored the \tech{}-refined tests.

Furthermore, we integrated \tech{} to improve ICL-based unit test generation approaches by refining the original test examples.
For this purpose, we selected the state-of-the-art \raggen{} and \telpa{} as the ICL-based approaches to be further improved. 
The former used \textit{developer-written} unit tests as examples while the latter employed \textit{tool-generated} ones as counter-examples, indicating diverse scenarios for evaluating \tech{}.
Our results demonstrate that \tech{}-refined test examples enable both \raggen{} and \telpa{} to achieve better effectiveness compared to using original examples or \utgen{}-refined examples. Specifically, we observe average improvements of 25.97\%, 28.22\%, and 45.99\% in CSR, PR, and Cov, respectively, when compared to \utgen{}-refined examples.
The ablation study reveals that both test purification and program analysis-based post-processing contribute significantly to \tech{}’s overall effectiveness.

To sum up, our work makes the following contributions:
\begin{itemize}
    \item We propose \tech{}, a novel test refinement technique to enhance the semantic clarity of unit tests by leveraging both LLMs and program analysis. We have also made the replication package publicly available~\cite{homepage}.

    \item We improve the effectiveness of ICL techniques for unit test generation from the novel perspective of enhancing the semantic clarity of in-context test examples. 

    \item We conducted an extensive study to evaluate \tech{} by measuring the quality (i.e., test-effectiveness-preserving degree and semantic clarity) of its refined unit tests and the improved effectiveness of ICL-based unit test generation with its refined unit test examples. 
    Additionally, we performed a user study to assess developers' perceptions of the refined tests, revealing their practical value not only for test generation but also for broader applications such as test maintenance and debugging. 
    
\end{itemize}

\section{Motivation}
\label{sec:motivation}

\lstdefinestyle{mystyle}{
    numbers=left,                   %
    numberstyle=\tiny\color{gray},   %
    basicstyle=\ttfamily\scriptsize,
    lineskip=0.5pt,                   %
    keywordstyle=\color{magenta},    %
    commentstyle=\color{violet!60!gray}, %
    stringstyle=\color{blue},         %
    backgroundcolor=\color{white}, %
    showstringspaces=false,          %
    xleftmargin=1.5em,               %
    xrightmargin=1.5em,              %
    frame=single,                    %
    breaklines=true,                 %
    moredelim=**[is][\color{red!50}]{`}{`}, %
    escapeinside=``,                 %
    columns==fullflexible,
}

\lstset{style=mystyle, aboveskip=0em}

We use a real-world example to motivate our work. 
Listing~\ref{listing:origin_example} shows an original unit test (that has been simplified for ease of illustration) for the method {\tt getColumnMatrix}, which checks two behaviors: (1) retrieving the column matrix at index 3, and (2) throwing an exception for index 5. However, its semantic clarity is poor.
\begin{lstlisting}[language=Java, caption={An example of an original unit test}, label={listing:origin_example}]
public void testGetColumnMatrix() {
    RealMatrix m = new RealMatrixImpl(subTestData);
    RealMatrix mColumn3 = new RealMatrixImpl(subColumn3);
    assertEquals("Column3", mColumn3, m.getColumnMatrix(3)); 
    assertThrows(MatrixIndexException.class, () -> m.getColumnMatrix(5));
}
\end{lstlisting}
First, it mixes two distinct scenarios (i.e.,valid and invalid index handling) within a single test. 
Mixing different scenarios within one test could aggravate the test complexity and thus negatively affect the semantic clarity to some degree. Second, ambiguous identifiers (e.g., {\tt mColumn3}) fail to convey their purpose, violating naming conventions and obscuring the test's intent.
Such unclear tests hinder LLMs from learning effectively about unit test generation when used as in-context examples. In fact, using this test in \raggen{} (an ICL-based unit test generation approach detailed in Section~\ref{sec:prompting}) yielded no improvement in line coverage compared to not using examples at all.

We then applied the state-of-the-art test refinement technique \utgen{}~\cite{utgen} to refine this unit test based on the DeepSeek-V2.5 model. 
As shown in Listing~\ref{listing:utgen_example}, the refined version improves textual clarity but unfortunately misinterprets the test's intent.
Specifically, the LLM treats the test as checking boundary cases rather than specific indices 3 and 5, evident from comments at Lines 5–6 and the use of {\tt matrix.getColumnDimension()-1} at Line 8, which causes an {\tt AssertionError} since the column at index 3 is not equal to the column at index {\tt matrix.getColumnDimension()-1}. Similarly, Line 13 incorrectly uses {\tt matrix.getColumnDimension()} instead of 5.
We investigated whether mixing different test scenarios contributes to this misunderstanding.
Specifically, we split the original test into two, each focusing on a single index, and refined them separately with \utgen{}.
The resulting tests correctly preserved the original intent, confirming that \textit{mixing scenarios likely caused the misinterpretation}.

Besides, such misinterpretation can aggravate LLM hallucinations and compromise the original unit tests' functionalities.
As shown in Listing~\ref{listing:utgen_example}, the \utgen{}-refined test calls a non-existent API {\tt getColumnDimension}, making it invalid. Using this flawed test as an example in \raggen{} led to lower line coverage (60.00\%) than using the original test (72.00\%).
This shows that \textit{poor refinement can even produce a negative effect on LLM-based test generation}.
Motivated by these challenges, we design a novel test refinement technique (called \tech{}) in this work, which aims to \textit{enhance the semantic clarity of unit tests while preserving effectiveness of the original test}, thereby enhancing ICL techniques for unit test generation by taking refined tests as in-context examples.

Note that while hallucinations in this case were addressed by splitting the original test and refining them separately with \utgen{}, this success stems from the simplicity of the example chosen for ease of illustration. 
In more complex scenarios, even on split tests, \utgen{} (solely relying on LLMs) still introduces errors due to hallucinations, such as API misuse, which has been confirmed by our ablation study in Section~\ref{sec:ablation}.
This underscores the need for a more robust approach like \tech{}, which \textit{combines test purification with program analysis to avoid errors caused by hallucinations and ensure both semantic clarity and functional correctness}.

\lstset{style=mystyle}
\begin{lstlisting}[language=Java, caption={The unit test refined by \utgen{}}, label={listing:utgen_example}, float=t, breaklines=true]
public void testRetrieveColumnMatrixWithValidAndInvalidIndices() {
    //Given: A RealMatrix initialized with subTestData
    RealMatrix matrix = new RealMatrixImpl(subTestData);
    RealMatrix expectedLastColumn = new RealMatrixImpl(subColumn3);
    //When: Retrieving the last column matrix
    //Then: The retrieved last column matrix should match the expected last column matrix
    `\colorbox{red!50}{assertEquals(expectedLastColumn, matrix.}`
    `\colorbox{red!50}{getColumnMatrix(matrix.getColumnDimension() - 1));}`
    //When: Attempting to retrieve a column matrix with an index equal to the column dimension
    //Then: A MatrixIndexException should be thrown
    `\colorbox{red!50}{assertThrows(MatrixIndexException.class,}`
        `\colorbox{red!50}{() -> matrix.getColumnMatrix(}`
        `\colorbox{red!50}{getColumnMatrix(matrix.getColumnDimension()));}`
}
\end{lstlisting}

\section{Methodology}
\label{sec:approach}
\begin{figure}[t]
    \centering
    \includegraphics[width=0.98\linewidth]{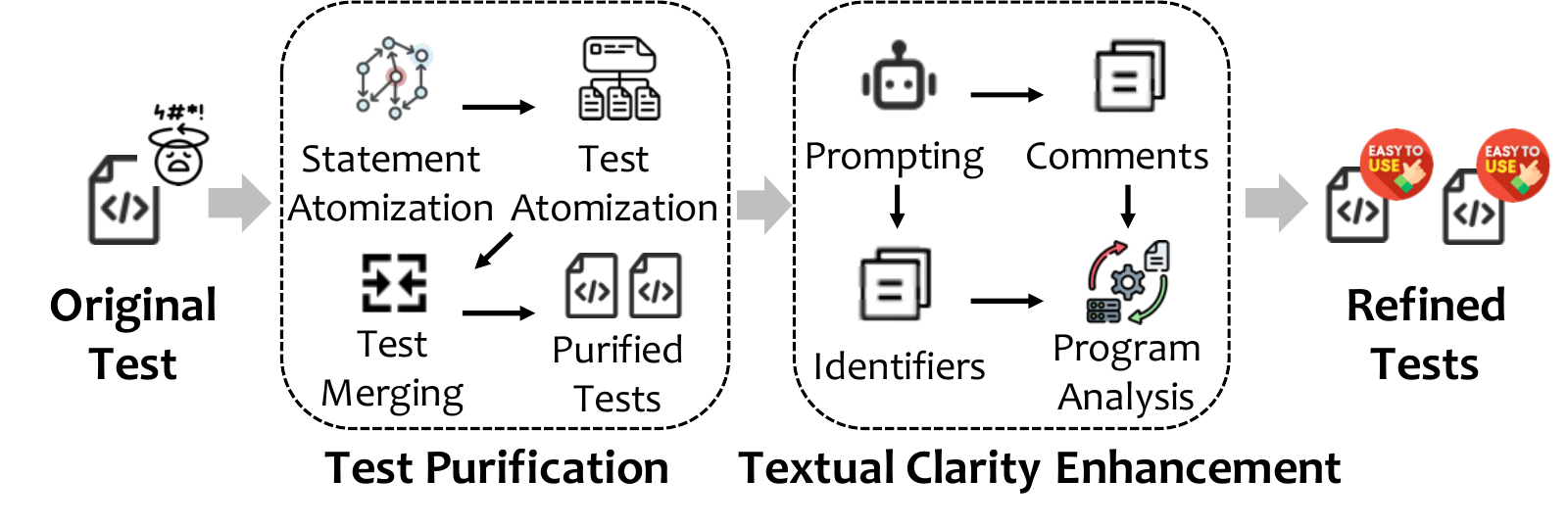}
\caption{Overview of \tech{}}
\label{fig:overview}

\end{figure}

Figure~\ref{fig:overview} provides an overview of \tech{}.
Given a (complex) unit test, \tech{} first applies a test purification component to decompose it into a set of simpler and purified tests. Next, \tech{} enhances the textual clarity of these purified tests using a program-analysis-enhanced approach.
This approach integrates program analysis with LLMs’ strong code comprehension capabilities to generate meaningful comments and more appropriate identifiers while minimizing the risk of errors caused by hallucinations.
As a result, \tech{} produces a set of refined tests that retain the original effectiveness while improving semantic expressiveness through clearer comments and identifiers, each targeting a single clear test scenario.
Then, these refined tests can be used as high-quality in-context examples in ICL-based unit test generation approaches, enabling LLMs to better learn effective patterns and thus improving the effectiveness of unit test generation.

\subsection{Test Purification}
\label{sec:slicing}
Test purification aims to produce a set of purified unit tests from each original test, with certain statements removed to isolate a single, clear test scenario.
While program slicing tools like Slicer4J~\cite{ahmed2021slicer4j} could theoretically support this process, they are ill-suited to our needs due to two main issues: (1) reliance on dynamic analysis, which requires compilation and execution to gather runtime traces, and (2) overly complex designs that introduce unnecessary overhead for simplifying small-scale unit tests. 
Therefore, \tech{} employs a lightweight static approach tailored for test purification, comprising three steps:
(1) Statement Atomization: breaking tests into atomic units to prevent syntax errors or unintended deletions in the next step;
(2) Test Atomization: splitting tests into simpler ones with a single assertion each. This is achieved through test slicing, which removes statements unrelated to the assertion, thereby simplifying complex logic.
(3) Test Merging: merging atomic tests with identical prefixes (indicating they target similar or identical scenarios) to reduce redundancy.

\subsubsection{Term Definition}
We first define some terms formally for ease of representation.
An \textbf{atomized statement} $S_a$ is a unit of code representing a single logical operation, either a standalone expression (e.g., variable declaration or method call) or a control structure (e.g., {\tt if}, {\tt for}, {\tt while}). 
Formally, $S_a = (T, V_r, V_w, C)$, where $T$ denotes the statement type (normal or control structure), $V_r$ is the set of variables read, $V_w$ is the set of variables written, and $C$ is a control flag ($C = \text{true}$ for control structures, i.e., $S_c$; otherwise, $S_n$). A variable is added to $V_r$ if it appears in $S_a$ and $S_a$ is not an assignment or declaration, or if it appears on the right-hand side of such a statement. It is added to $V_w$ if it appears on the left-hand side of an assignment/declaration, or if it serves as the caller or parameter in a method call whose name implies modification (e.g., ``set'', ``add'', ``insert'', ``remove''). A full keyword list is available on our project homepage~\cite{homepage}.

\subsubsection{Statement Atomization} 
A single statement may contain multiple operations (e.g., {\tt int a, b;} or {\tt a = b = 1;}). To avoid unintended deletions or syntax errors when removing statements (at the next step), \tech{} first breaks compound statements into atomic components. For a compound statement $S$, it produces a set of atomized statements $\{S_{a1}, S_{a2}, \dots, S_{ak}\}$, each in the form $(T_i, V_{ri}, V_{wi}, C_i)$. 
Specifically, variable declarations with multiple identifiers are split into separate statements based on the type and identifiers; chained assignments are similarly decomposed. 
\ins{
\tech{} handles only multiple declarations and chain assignments, leaving other instructions (e.g., method call chains) unchanged to avoid unintended deletions later. This conservative strategy balances test-effectiveness preservation with brevity.
}
Control structures are treated as indivisible units (e.g., $S_c = (T_c, V_{rc}, V_{wc}, \text{true})$) to avoid syntax issues such as removing only a {\tt for} loop's condition expression, which would result in an incomplete loop.
The body of the control structure is treated as a separate set of statements. This atomization reduces the risk of producing errors during subsequent test atomization.

\subsubsection{Test Atomization}
A unit test comprises two parts: \textit{test prefix} and \textit{assertion}~\cite{chattester}. 
The test prefix sets up the focal method with a series of method calls or assignments, while the assertion verifies its expected behavior.
A test may contain multiple assertions, potentially testing different scenarios. 
To enhance simplicity and clarity, \tech{} atomizes the test by splitting it into multiple tests, each containing only one assertion, and then slices each test to remove unrelated statements.

Formally, let the prefix $T_p$ be a sequence of atomized statements $\{S_{a1}, S_{a2}, \dots, S_{am}\}$, and $A = \{a_1, a_2, \dots, a_n\}$ be the set of assertions. For each assertion $a_i$, \tech{} creates a new test $T_i = T_p + a_i$.
Then, \tech{} performs backward slicing on each test \( T_i \) to remove statements unrelated to \( a_i \). 
Specifically, \tech{} builds a variable dependency graph $G = \{V, E\}$, where $V$ are variables and $E$ are directed edges $e_{v_i \to v_j}$ showing dependencies. Starting from each variable used in $a_i$, it collects all its reachable variables in $G$ as the variables it depends on (denoted as $V_{\textit{depen}}$). 
Then, for each normal statement \( S_{\textit{nj}} \in T_p \), \tech{} removes \( S_{\textit{nj}} \) if \( V_{\textit{wj}} \cap V_{\textit{depen}} = \emptyset \). This ensures that only statements contributing to the assertion \(a_i\) are retained.
Finally, any control structure $S_{\textit{cj}}$ with an empty body is removed.

\subsubsection{Test Merging}
After completing the above steps, we obtain tests each with a single assertion and its relevant prefix. 
However, some tests may share identical prefixes and validate different aspects of the same behavior through different assertions. 
To reduce redundancy, \tech{} merges the group of tests that share the same prefix into a single test by combining the shared prefix and all the assertions within these tests.
Formally, given tests $\{T_1, T_2, \dots, T_k\}$ sharing prefix $T_p$, \tech{} merges them into $T_{\text{merged}} = T_p + \{a_1, a_2, \dots, a_k\}$, where $a_i$ is the assertion from $T_i$.
Note that while some highly-focused tests may be merged back into their original form, in most cases, the output for an original test is a set of logically clearer tests.

\subsection{Textual Clarity Enhancement}
Many studies have emphasized the importance of clear comments and meaningful identifiers for semantic clarity~\cite{testdescriber, Daka, deeptc}.
Hence, the goal of this component is to enhance semantic expressiveness of a given unit test by refining the two kinds of elements within the unit test. 
This process begins by using an LLM to generate comments or identifiers through carefully-crafted prompts, 
and then employs program analysis to integrate the LLM response with the original unit test, ensuring the preservation of test effectiveness and avoiding potential errors from LLM's hallucination. 
Note that the key contribution is not the idea of generating comments or identifiers with LLMs, which has been demonstrated by previous research~\cite{deeptc,utgen}.
Rather, it is the program-analysis-based post-processing that ensures refinement accuracy and mitigates hallucination issues.

\subsubsection{Comment and Identifier Generation via LLM Prompting}
\tech{} employs in-context learning to construct prompts by providing task-specific examples and instructions to guide the LLM.
Specifically, \tech{} applies one-shot in-context learning, which uses a single high-quality example to enhance the LLM's comprehension. 
For comment generation, the example includes a unit test alongside its enhanced version with comments following the ``Arrange-Act-Assert'' pattern~\cite{wei2023automatically}. 
For identifier generation, the example includes the original identifiers from the test alongside their enhanced version with carefully crafted expressive names.
Our prompt design draws on the best practices derived from recent advances in prompt engineering research.
Due to the space limit, the prompts used for generating comments and identifiers in \tech{} are presented on our project homepage~\cite{homepage}.

\subsubsection{Post Processing via Program Analysis}
In this process, \tech{} extracts comments and identifiers from the above LLM-generated contents, and then employs program analysis, especially AST node matching, to seamlessly integrate them with the original unit tests.

For comment post-processing, \tech{} first parses the AST of the LLM-generated test to extract block and inline comments. 
Block comments, typically docstrings, are placed at the beginning of the test. 
For inline comments, since LLM-refined tests may alter statements due to hallucinations, \tech{} must map these comments back to the original test.
Therefore, \tech{} extracts their immediate right sibling nodes as context. Consecutive inline comments are merged into a single node before context extraction.
Then, \tech{} traverses the original test’s AST, comparing each statement node to the comment’s context node using syntactic and semantic similarity following prior work~\cite{gao2019teccd, oh2024csa}.
Formally, for nodes $v_1$ (in original test) and $v_2$ (in refined test), the similarity is defined as:
$$
\textit{distance} = \textit{type\_match}(v_1, v_2) \times \textit{CodeBLEU}(v_1, v_2)
$$
Here, \textit{type\_match} checks whether the node types of the two nodes are consistent (returning 1 for a match and 0 for a mismatch), while \textit{CodeBLEU} represents the CodeBLEU similarity~\cite{codebleu} between the nodes. 
If the similarity score exceeds a certain threshold, \tech{} considers it a match and inserts the inline comment before the corresponding statement.
Here, we conducted a preliminary study on a small test benchmark and then set the threshold to 0.7 based on the observed results.

For identifier post-processing, \tech{} first extracts all the original identifiers and corresponding new identifiers from the LLM's output.
Note that if the identifiers generated by the LLM contain duplicates, we re-instruct the LLM to avoid them.
It then traverses the AST of the original unit test to extract all variable declaration nodes, and creates a mapping of these declaration nodes with their associated identifiers. 
Next, \tech{} traverses the AST to identify all variable identifiers and record their positions. In reverse order, it replaces each identifier with the newly generated name suggested by the LLM, ensuring that replacements do not shift subsequent positions and cause errors.
Finally, for the test name, \tech{} locates the method declaration of the unit test and replaces the original test name with the newly-generated name.

\lstdefinestyle{mystyle}{
    numbers=left,                   %
    numberstyle=\tiny\color{gray},   %
    basicstyle=\ttfamily\scriptsize, %
    lineskip=0.5pt,                   %
    keywordstyle=\color{magenta},    %
    commentstyle=\color{violet!60!gray}, %
    stringstyle=\color{blue},         %
    backgroundcolor=\color{white}, %
    showstringspaces=false,          %
    xleftmargin=1.5em,               %
    xrightmargin=1.5em,              %
    frame=single,                    %
    breaklines=true,                 %
    moredelim=**[is][\color{red!50}]{`}{`}, %
    escapeinside=``,                 %
}

\lstset{style=mystyle, aboveskip=0em, belowskip=-1.9em}

\begin{lstlisting}[language=Java, caption={The unit test refined by \tech{} (docstrings are omitted here for brevity)}, label={listing:our_example}, float=t]
/* Omitted for saving space */
public void testRetrieveColumnAsSubMatrix() {
    // Arrange: Create a RealMatrix instance using subTestData
    RealMatrix matrixUnderTest = new RealMatrixImpl(subTestData);
    // Create a RealMatrix instance representing the expected column matrix for column 3
    RealMatrix expectedColumnMatrix = new RealMatrixImpl(subColumn3);
    `\colorbox{green!50}{// Act: Retrieve the column matrix at index 3}`
    `\colorbox{green!50}{RealMatrix retrievedColumnMatrix=matrixUnderTest}`
                    `\colorbox{green!50}{.getColumnMatrix(3);}`
    // Assert
    // Verify that the retrieved column matrix matches the expected column matrix
    assertEquals(expectedColumnMatrix, retrievedColumnMatrix);
}
/* Omitted for saving space */
public void 
testGetColumnMatrixWithInvalidIndicesThrowsException() 
{
    // Arrange
    RealMatrix matrixInstance = new RealMatrixImpl(subTestData);
    // Act and Assert
    `\colorbox{green!50}{// Get a column matrix with an index out of bounds}`
    `\colorbox{green!50}{assertThrows(MatrixIndexException.class,}`
        `\colorbox{green!50}{() -> matrixInstance.getColumnMatrix(5));}`
}
\end{lstlisting}

In this way, \tech{} outputs one or more refined tests with clearer logic, better comments, and more descriptive identifiers, for a given original test. These are then used as in-context examples to boost ICL-based test generation. When multiple tests are produced, all are included to preserve completeness and ensure fidelity to the original test.
Listing~\ref{listing:our_example} shows how \tech{} splits mixed scenarios into two focused tests (valid and invalid index cases),enhancing clarity and intent. For example, Line 7’s comment precisely describes the action at Line 8. Using these refined tests in \raggen{} raised coverage to 84.00\%, versus 72.00\% with the original test example and 60.00\% with \utgen-refined test example.

\section{Evaluation Design}
\subsection{Research Questions}
\smallskip
\noindent
\textbf{RQ1: How effective is \tech{} in refining unit tests?} 
We evaluated whether \tech{} preserves the effectiveness (e.g., compilation success rate, pass rate, test coverage) of original unit tests while improving their semantic clarity.

\smallskip
\noindent
\textbf{RQ2: To what extent do the refined unit tests by \tech{} enhance the effectiveness of ICL-based unit test generation?}
We assessed whether \tech{}-refined examples improve the effectiveness of state-of-the-art ICL-based approaches (i.e., \raggen{}~\cite{ase_empirical}, \telpa{}~\cite{telpa}) compared to using original or baseline-refined examples.

\smallskip
\noindent
\textbf{RQ3: How does each component in \tech{} contribute to its effectiveness?}
We conducted an ablation study to evaluate the impact of \tech{}'s key components: test purification and program-analysis-based post-processing.

\subsection{Subjects}
Following the existing work in unit test generation~\cite{chatunitest,ase_empirical}, we evaluated \tech{} on Java projects using the JUnit framework~\cite{junit} due to the significant popularity of Java and JUnit.
In total, we used seven real-world Java projects as our subjects, including four open-source projects and three industrial projects.
Following the existing work~\cite{coderujb}, we used the latest versions of four Java projects in the widely-used Defects4J benchmark~\cite{defects4j} (i.e., Chart, Time, Lang, and Math), excluding Closure due to its lack of JUnit tests and high testing costs.
While these projects may appear in the LLM training data, the models were not specifically trained for test generation or refinement, which mitigates potential data leakage concerns.

Particularly, to further reduce data leakage risks, we adopted three internal Java projects provided by our industrial partner (a global leader company in IT).
The three industrial projects have different functionalities, i.e., a program analysis toolkit, an online micro-service system, and a data analysis framework involving parallel computing and adaptation of design patterns. For ease of presentation, we refer to them as PATool, Microservice and DAService in the following sections.
Due to the company policy, we are unable to disclose further details.
This diverse range of subjects allows us to thoroughly evaluate \tech{}'s generalizability.

\subsection{Metrics}
\label{sec:metrics}
We first evaluated refined unit tests using two key measurements: \textit{test-effectiveness-preserving degree} and \textit{semantic clarity}.
In line with the existing studies~\cite{utgen}, we measured whether refinement preserves test effectiveness using three widely-used metrics: Compilation Success Rate (CSR), Pass Rate (PR), Line Coverage (Cov)\ins{, and Mutation Score (MS)}. CSR and PR represent the ratios of successfully compiled and executed tests, respectively, Cov measures the average line coverage across focal methods, and \ins{MS measures the proportion of artificially injected faults (mutants) that the tests successfully detect~\cite{chattester, symprompt, chen2023toward}.}
We calculated the difference between refined and original tests for each metric. Zero differences indicate full effectiveness preservation, while negative values suggest effectiveness degradation. 

To evaluate the semantic clarity of refined tests, following prior work~\cite{deeptc,utgen,Daka}, we conducted a user study with 15 participants experienced in Java and testing, averaging 6.8 years of development experience (10 from industry and 5 from academia). 
\ins{We randomly sampled 10 focal methods from the four Defects4J projects ($\ge$ 1 per project) for diversity}
and asked participants to rank the original test and the tests refined by different techniques for each method on three criteria: (1) \textit{conciseness} (clarity of test scenarios), (2) \textit{descriptiveness} (quality of identifiers), and (3) \textit{comment quality} (clarity of comments).
\ins{
We adopted a rank-based scale to minimize rating bias and support clearer comparisons. 
}
A rank of 1$^{\textit{st}}$ indicates the best performance. To avoid bias, participants were unaware of which technique was used for each test.

We then evaluated the impact of refined tests on ICL-based test generation approaches by measuring CSR, PR, and Cov of the generated tests by these approaches using in-context examples from different refinement techniques.

\subsection{Studied ICL-based Unit Test Generation Approaches}
\label{sec:prompting}

To evaluate \tech{}’s impact on ICL-based test generation, we enhanced two state-of-the-art approaches (i.e., \raggen{}~\cite{ase_empirical} and \telpa{}~\cite{telpa}). They incorporated different types of unit tests as in-context examples with different purposes, indicating diverse scenarios for evaluating \tech{}.

\raggen{} retrieves the most similar method to the focal method and its \textit{developer-written} unit tests from a database.
Then, it incorporates the identified method and all its associated unit tests as examples in the prompt, which enables LLMs to learn more knowledge on generating unit tests.
\telpa{} first uses the widely-used search-based tool (i.e., EvoSuite~\cite{evosuite}) to generate initial tests, then switches to LLMs when coverage is insufficient, using \textit{tool-generated} tests as counter-examples.

For more sufficient evaluation, we used two LLMs for both approaches: CodeLlama-7b-Instruct-hf (\textit{CL-7B}) and deepseek-coder-6.7b-instruct (\textit{DS-7B}).
Both LLMs have been demonstrated effective in code-related tasks~\cite{huggingface_leaderboard,codellama,deepseek_paper}, and the $\sim$7b size balances cost and effectiveness well.
Note that our goal is to evaluate whether refining in-context tests examples can enhance the effectiveness of \raggen{} and \telpa{}, rather than compare the effectiveness of different LLMs, and thus we only need to control the same LLM when comparing refined and original unit tests on an ICL-based approach.
Studying two underlying LLMs for each ICL-based approach helps improve the evaluation's generalizability.

\subsection{Baselines}
First, we should understand the quality difference between unit tests refined by \tech{} and the original unit tests without refinement, thus we treated the original unit tests (denoted as \textbf{\origin{}}) as one baseline.
Then, we should understand the effectiveness of \tech{} compared to other test refinement techniques.
Here, we selected the state-of-the-art unit test refinement technique (i.e., \textbf{\utgen{}}~\cite{utgen}) as another baseline.
It employs an LLM to contextualize test data and improve textual clarity.

\subsection{Implementation and Environment}
We implemented \tech{} in Python, using the tree-sitter tool~\cite{tree-sitter} for AST analysis and DeepSeek-V2.5 via its API~\cite{deepseek} as the underlying LLM. For open-source projects, we used Defects4J's framework and Cobertura~\cite{cobertura} for coverage collection. Industrial projects were tested using internal frameworks. Experimental scripts were developed with PyTorch 2.0.0~\cite{torch} and Transformers 4.34.1~\cite{transformers}, accelerated by the vLLM library~\cite{vllm}. For \utgen{}, \telpa{}, and \raggen{}, we used their publicly released artifacts. 
For fair comparisons between \tech{} and \utgen{}, we also employed DeepSeek-V2.5 as the underlying LLM for \utgen{}.
All experiments ran on an Ubuntu 18.04 server with an Intel Xeon Gold 6240C CPU, 512GB RAM, and four NVIDIA A800 GPUs.

\section{Results and Analysis}

\subsection{RQ1: Effectiveness Comparison in Unit Test Refinement}

\subsubsection{Process}
In this RQ, we examined the quality of the refined unit tests by \tech{}.
We considered two types of unit tests: \textbf{developer-written tests} and \textbf{tool-generated tests}.
For developer-written tests, we gathered all unit tests included in each open-source project.
For tool-generated tests, we employed EvoSuite~\cite{evosuite}, a widely-used search-based test generation tool, to automatically create a test suite for each project.
Then, we sampled 500 developer-written unit tests and 500 tool-generated unit tests for refinement by balancing generalizability and cost similar to the existing work~\cite{chattester, coderujb}.
Note that we excluded the industrial projects in this RQ for two reasons. First, the developer-written unit tests for these projects are not accessible. Second, the three industrial projects use Java 17, but EvoSuite only supports up to Java 11, making it impossible to generate tests for them.

We then used \utgen{} and \tech{} to refine the two sets of original unit tests (directly used by the \origin{} baseline). Subsequently, we measured the quality of the original, \utgen{}-refined, and \tech{}-refined unit tests. We assessed their test-effectiveness-preserving degree using CSR, PR, and Cov metrics. Additionally, we evaluated semantic clarity in terms of conciseness, descriptiveness, and comment quality through a user study.

\subsubsection{Results}

\begin{table*}[t]
\caption{Comparison between \tech{}, Origin, and \utgen{} in test-effectiveness-preserving degree (RQ1)}
\vspace{-0.6em}
\large
\centering
\label{tab:rq1_function}
\renewcommand{\arraystretch}{1}
\resizebox{0.65\linewidth}{!}{
\begin{threeparttable}
\begin{tabular}{l|rrrr|rrrr}
\toprule
\multirow{2}{*}{\textbf{Technique}} & \multicolumn{4}{c|}{\textbf{Developer-written Tests}} & \multicolumn{4}{c}{\textbf{Tool-generated Tests}} \\
                               & \multicolumn{1}{c}{\textbf{CSR}} & \multicolumn{1}{c}{\textbf{PR}} & \multicolumn{1}{c}{\textbf{Cov}} & \multicolumn{1}{c|}{\textbf{MS}} & \multicolumn{1}{c}{\textbf{CSR}} & \multicolumn{1}{c}{\textbf{PR}} & \multicolumn{1}{c}{\textbf{Cov}} & \multicolumn{1}{c}{\textbf{MS}} \\
\midrule
Origin                & 100.00\%    & 99.75\%    & 48.49\%  & 73.97\%   & 100.00\%    & 100.00\%    & 43.43\%   & 57.21\%  \\
$\Delta$\utgen{}                 & -10.07\%    & -32.30\%   & -3.76\% & -3.29\%   & -15.73\%    & -39.33\%    & -5.53\%   & -6.84\%   \\
$\Delta$\tech{}                  & 0.00\%      & 0.00\%  & 0.00\%    & 0.00\%       & 0.00\%      & 0.00\%      & 0.00\%   & 0.00\%      \\
\bottomrule
\end{tabular}
\end{threeparttable}
}
\end{table*}

\begin{figure}[t]
\setlength{\abovecaptionskip}{-0.02cm} 
    \centering
    \includegraphics[width=0.86\linewidth]{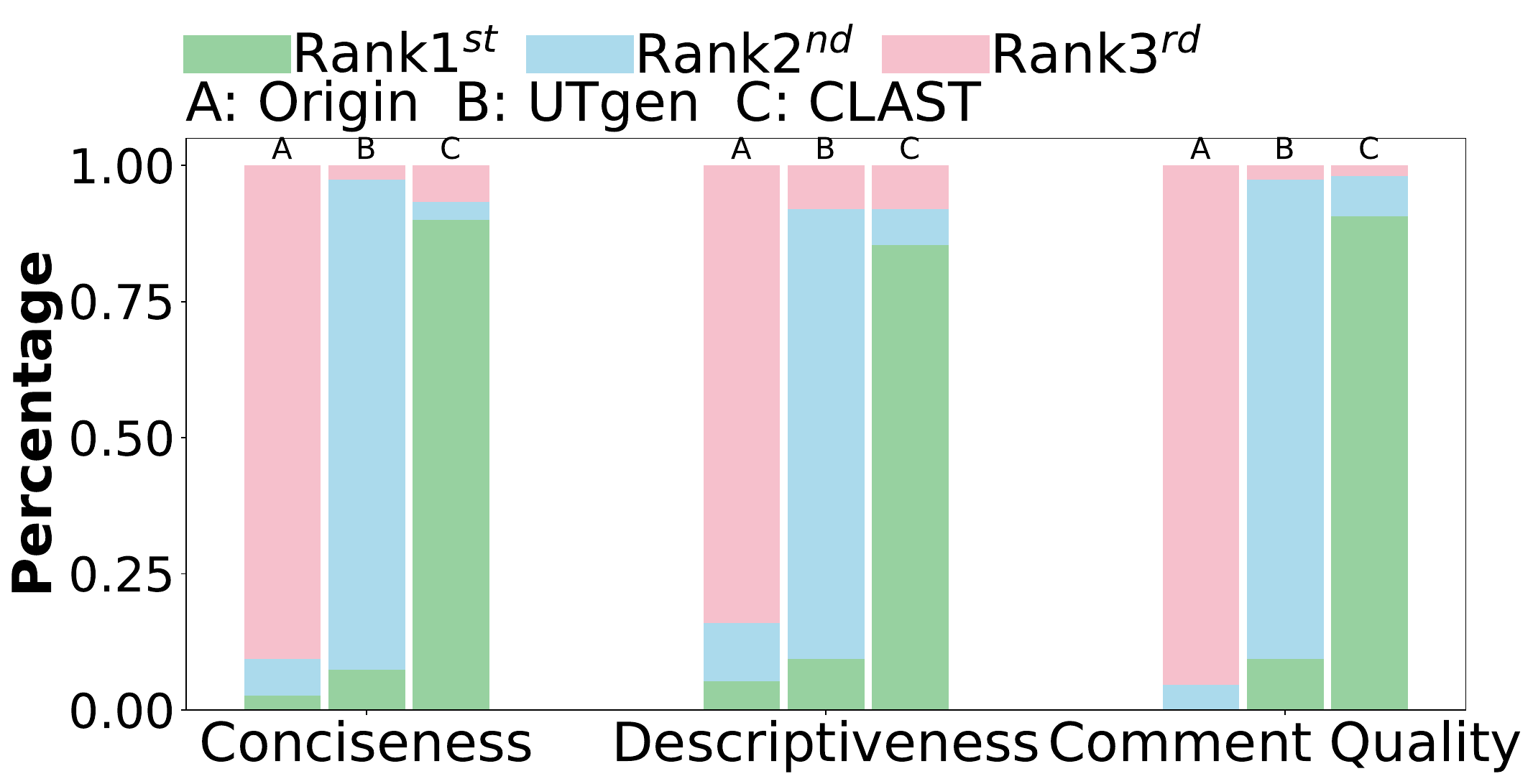}
\caption{Comparison between \tech{}, Origin, and \utgen{} in terms of semantic clarity (RQ1)}
\label{fig:rq1_user}
\vspace{-1em}
\end{figure}

Table~\ref{tab:rq1_function} presents the comparison results for the test-effectiveness-preserving degree, where Rows ``$\Delta$\utgen{}'' and ``$\Delta$\tech{}'' represent the difference between refined and original unit tests in terms of the corresponding metric.
From this table, \tech{} maintains unit test effectiveness after refinement, while \utgen{} experiences significant declines across all three metrics for both developer-written and tool-generated tests. 
For example, for tool-generated tests, \tech{} achieves identical CSR (100.00\%), PR (100.00\%), Cov (43.43\%)\ins{, and MS(57.21\%)} compared to \origin{}, while \utgen{}'s performance deteriorates, with CSR, PR, Cov, and MS decreases by 15.73\%, 39.33\%, 5.53\%, \ins{and 6.84\%} respectively.
This decline likely stems from the LLM's hallucination problem.

Figure~\ref{fig:rq1_user} shows the comparison results for the semantic clarity.
In this figure, each bar shows the ranking distribution for the tests refined by a technique in terms of a metric.
The user study demonstrates that \tech{}-refined tests exhibit better semantic clarity, improving their readability and comprehensibility.
Specifically, \tech{}-refined unit tests receive the highest percentage of first-place rankings across all the three metrics compared to those refined by \utgen{} and the original tests: 90.00\% for conciseness, 85.33\% for descriptiveness, and 90.67\% for comment quality.
Note that among the ten randomly sampled methods for the user study, five are with developer-written tests and the other five are with tool-generated tests. The conclusions remain consistent across both types.
This highlights participants' strong preference for \tech{}-refined unit tests, validating the practical value of \tech{} in enhancing semantic clarity.

\begin{tcolorbox}\textbf{RQ1 Summary:}
\tech{} exhibits the superior test-effectiveness-preserving ability while significantly enhancing test semantic clarity compared to \utgen{}.
Specifically, \tech{} completely retains the original effectiveness of the unit tests, whereas \utgen{} decreases CSR, PR, Cov, and \ins{MS} by \ins{12.90}\%, \ins{35.82}\%, \ins{4.65}\%, and \ins{5.07\%} on average. 
Furthermore, over 85.33\% of the user study participants favored the semantic clarity (i.e., conciseness, descriptiveness, and comment quality) of the unit tests refined by \tech{}.
\end{tcolorbox}

\subsection{RQ2: Effectiveness Comparison in Enhancing Unit Test Generation}
\label{sec:results_rq2}

\subsubsection{Process}
For each studied ICL-based test generation approach (i.e., \raggen{} and \telpa{}) with each studied LLM (i.e. CL-7B and DS-7B), we incorporated the original, \utgen{}-refined, \tech{}-refined unit test examples into the prompt for generating unit tests, respectively.
To better understand the effect of incorporating in-context test examples, we also ran these approaches without any test examples (denoted as ``Base'' for ease of presentation).
Then, we measured the effectiveness of the generated unit tests by each technique in terms of CSR, PR, and Cov.
As previously mentioned, the three industrial projects are based on Java 17 but the EvoSuite tool used by \telpa{} only supports up to Java 11, and thus we just applied \telpa{} to the four open-source projects in this experiment.

\begin{table*}[t]
\caption{Comparison between \tech{}, Origin, and \utgen{} in terms of the effectiveness for test generation (RQ2) }
\vspace{-0.7em}
\centering
\label{tab:rq2_function}
\renewcommand{\arraystretch}{1.2}
\resizebox{0.8\linewidth}{!}{
\begin{threeparttable}
\begin{tabular}{l|l|l|rrrr|rrrr}
\toprule
\multirow{2}{*}{\textbf{App.}} & \multirow{2}{*}{\textbf{Metric}} & \multirow{2}{*}{\textbf{Project}} & \multicolumn{4}{c|}{\textbf{CL-7B}} & \multicolumn{4}{c}{\textbf{DS-7B}} \\
                               &                                  &                                   & \multicolumn{1}{l}{\textbf{Base}} & \multicolumn{1}{l}{\textbf{Origin}} & \multicolumn{1}{l}{\textbf{\utgen{}}} & \multicolumn{1}{l|}{\textbf{\tech{}}} & \multicolumn{1}{l}{\textbf{Base}} & \multicolumn{1}{l}{\textbf{Origin}} & \multicolumn{1}{l}{\textbf{\utgen{}}} & \multicolumn{1}{l}{\textbf{\tech{}}} \\
\midrule
\multirow{18}{*}{\raggen{}}          & \multirow{7}{*}{CSR}             & Time                              & 41.02\% & 55.66\% & 44.12\% & \textbf{65.00\%} & 59.26\% & 66.03\% & 27.64\% & \textbf{67.96\%} \\
                               &                                  & Math                              & 44.51\% & 41.86\% & 22.22\% & \textbf{66.67\%} & 51.63\% & 53.85\% & 48.15\% & \textbf{54.41\%} \\
                               &                                  & Lang                              & 33.70\% & 55.75\% & 62.55\% & \textbf{72.14\%} & 75.70\% & 76.84\% & 51.19\% & \textbf{87.54\%} \\
                               &                                  & Chart                             & 50.45\% & 63.13\% & 46.30\% & \textbf{64.38\%} & 70.00\% & 79.07\% & 52.03\% & \textbf{79.12\%} \\
                               &                                  & MicroService                      & 35.71\% & \textbf{42.86\%} & 40.00\% & \textbf{42.86\%} & \textbf{38.46\%} & 23.08\% & 15.38\% & \textbf{38.46\%} \\
                               &                                  & PATool                         & 82.86\% & 77.14\% & 82.86\% & \textbf{88.57\%} & 85.71\% & 82.86\% & 71.43\% & \textbf{91.43\%} \\
                               &                                  & DAService                         & 41.18\% & 44.12\% & 41.18\% & \textbf{52.94\%} & 31.43\% & 37.14\% & 37.14\% & \textbf{48.57\%} \\
\cline{2-11}
                               & \multirow{7}{*}{PR}              & Time                              & 22.09\% & 33.03\% & 31.15\% & \textbf{34.30\%} & 37.04\% & 42.31\% & 6.54\% & \textbf{45.63\%} \\
                               &                                  & Math                              & 26.40\% & \textbf{30.23\%} & 22.58\% & 28.42\% & 30.59\% & 34.62\% & 39.58\% & \textbf{39.71\%} \\
                               &                                  & Lang                              & 16.87\% & 22.71\% & 31.85\% & \textbf{32.30\%} & 30.39\% & 24.29\% & 25.49\% & \textbf{50.33\%} \\
                               &                                  & Chart                             & 22.97\% & 36.60\% & 24.05\% & \textbf{41.37\%} & 35.71\% & 42.44\% & 32.12\% & \textbf{53.82\%} \\
                               &                                  & MicroService & 18.73\% & 29.82\% & 15.69\% & \textbf{32.31\%} & 25.34\% & 27.87\% & 18.49\% & \textbf{26.43\%} \\
                               &                                  & PATool    & 40.76\% & 34.72\% & 38.58\% & \textbf{49.92\%} & 47.36\% & 51.42\% & 49.13\% & \textbf{54.98\%} \\
                               &                                  & DAService        & 22.14\% & 27.40\% & 25.35\% & \textbf{30.74\%} & 33.73\% & 35.43\% & 37.29\% & \textbf{47.42\%} \\
\cline{2-11}
                               & \multirow{4}{*}{Cov}             & Time         &                    51.92\%  & 57.32\% &  22.94\% & \textbf{63.41\%} & 43.42\% & 56.63\% & 18.12\% & \textbf{70.06\%} \\
                               &                                  & Math                            & 45.57\%  & 52.43\%  & 14.52\% & \textbf{62.97\%} & 49.41\% & 51.88\% & 42.44\% & \textbf{61.42\%} \\
                               &                                  & Lang                            & 33.10\%  & 63.22\%  & 50.31\% & \textbf{71.80\%} & 55.95\% & 71.56\% & 25.39\% & \textbf{79.10\%} \\
                               &                                  & Chart                           & 28.54\%  & 59.03\%  & 43.59\% & \textbf{63.31\%} & 48.96\% & 60.01\% & 27.04\% & \textbf{70.52\%} \\
                               &                                  & MicroService                      & 19.64\% & 29.21\% & 30.79\% & \textbf{31.29\%} & 20.46\% & 16.54\% & 11.54\% & \textbf{29.54\%} \\
                               &                                  & PATool                         & 72.03\% & 66.11\% & 65.03\% & \textbf{76.06\%} & 71.86\% & 70.26\% & 60.31\% & \textbf{78.91\%} \\
                               &                                  & DAService                         & 23.06\% & 26.03\% & 22.97\% & \textbf{30.38\%} & 24.60\% & 23.60\% & 25.00\% & \textbf{35.49\%} \\
\midrule
\multirow{12}{*}{\telpa{}}        & \multirow{4}{*}{CSR}             & Time                              & 68.77\% & 70.00\% & 59.26\% & \textbf{77.33\%} & 41.38\% & 67.01\% & 66.34\% & \textbf{69.91\%} \\
                               &                                  & Math                              & 30.00\% & 71.43\% & \textbf{80.00\%} & \textbf{80.00\%} & 33.33\% & \textbf{88.89\%} & \textbf{88.89\%} & \textbf{88.89\%} \\
                               &                                  & Lang                              & 81.79\% & \textbf{92.63\%} & 86.96\% & 92.55\% & 69.57\% & 93.59\% & 91.04\% & \textbf{94.17\%} \\
                               &                                  & Chart                             & 72.49\% & 83.02\% & 80.00\% & \textbf{94.12\%} & 78.75\% & 89.47\% & 84.31\% & \textbf{90.29\%} \\
\cline{2-11}
                               & \multirow{4}{*}{PR}              & Time                              & 17.89\% & \textbf{58.00\%} & 37.04\% & 57.33\% & 25.86\% & 42.27\% & 40.59\% & \textbf{43.36\%} \\
                               &                                  & Math                              & 20.00\% & 57.14\% & \textbf{60.00\%} & \textbf{60.00\%} & 0.00\% & \textbf{55.56\%} & \textbf{55.56\%} & \textbf{55.56\%} \\
                               &                                  & Lang                              & 34.97\% & \textbf{71.05\%} & 55.43\% & 70.21\% & 36.96\% & 63.46\% & 58.96\% & \textbf{65.83\%} \\
                               &                                  & Chart                             & 22.61\% & 33.96\% & 50.77\% & \textbf{60.50\%} & 41.25\% & 52.15\% & 52.18\% & \textbf{56.00\%} \\
\cline{2-11}
                               & \multirow{4}{*}{Cov}             & Time         &                    21.75\%  & 36.64\% &  32.19\% & \textbf{41.78\%} & 15.78\% & 28.05\% & 30.60\% & \textbf{33.77\%} \\
                               &                                  & Math                            & 29.17\%  & 31.13\%  & \textbf{42.16\%} & \textbf{42.16\%} & 17.95\% & 40.72\% & \textbf{44.99\%} & \textbf{44.99\%} \\
                               &                                  & Lang                            & 20.59\%  & 23.08\%  & 26.40\% & \textbf{35.20\%} & 18.87\% & 32.81\% & 34.43\% & \textbf{37.86\%} \\
                               &                                  & Chart                           & 17.92\%  & 19.78\%  & 28.03\% & \textbf{30.94\%} & 18.37\% & 25.00\% & 25.76\% & \textbf{26.31\%} \\
\bottomrule
\end{tabular}
\end{threeparttable}
}
\vspace{-1em}
\end{table*}

\subsubsection{Results}

Table~\ref{tab:rq2_function} compares the effectiveness of generated unit tests.
In this table, Columns ``Base'' represent the results of these ICL-based unit test generation approaches without test examples, while ``\origin{}'', ``\utgen{}'', and ``\tech{}'' represent the results of these ICL-based approaches with original, \utgen{}-refined, and \tech{}-refined examples, respectively.
The best results among the four techniques (Base, \origin{}, \utgen{}, and \tech{}) are marked as bold on each project in terms of each metric.
Note that \raggen{} targets all focal methods within a project while \telpa{} just targets the focal methods that are not fully covered by the used search-based tool within the given testing period, and thus it is meaningless to compare \raggen{} and \telpa{} based on this table.

From Table~\ref{tab:rq2_function}, we first compare the effectiveness of \origin{} and Base to investigate the effect of incorporating test examples into the prompt. 
In most cases, \origin{} indeed helps generate more effective unit tests than Base. 
For example, \origin{} improves the effectiveness of \raggen{} with CL-7B and DS-7B by 29.03\% and 11.38\% compared to Base in terms of the average line coverage across all the studied projects.
This demonstrates the value of incorporating unit test examples into the prompt for LLM-based unit test generation.
However, the improvement of \origin{} over Base is limited in many cases, e.g., Cov is just improved from 49.41\% to 51.88\% for \raggen{} with DS-7B on Math.
This may be attributed to the lack of semantic clarity for the test examples, thereby preventing the LLM from effectively learning from them.

We then compare the effectiveness of \tech{}, \origin{}, and \utgen{}, for enhancing ICL-based unit test generation.
From Table~\ref{tab:rq2_function}, \tech{} almost always performs the best among them.
For instance, on average across all the studied projects, using \raggen{} with CL-7B, \tech{} improves the Base, \origin{}, and \utgen{} methods by 37.38\%, 18.93\%, and 33.41\% in CSR; 46.72\%, 16.25\%, and 31.76\% in PR; and 45.78\%, 12.98\%, and 59.59\% in Cov, respectively.
For \raggen{} with DS-7B, \tech{} enhances the Base, \origin{}, and \utgen{} methods by 13.42\%, 11.61\%, and 54.31\% in CSR; 32.54\%, 23.20\%, and 52.57\% in PR; and 35.08\%, 21.27\%, and 102.55\% in Cov, respectively.
These trends persist for both \telpa{} with CL-7B and \telpa{} with DS-7B.
In the few instances where \tech{} is not the top performer, these occurrences are limited to CSR and PR measurements with only slight underperformance. However, the corresponding Cov achieved by \tech{} is significantly higher than those by baselines, potentially offering greater practical value.

Besides, we find that \utgen{}-refined unit tests often produce a negative influence on ICL-based unit test generation compared to directly using the original tests.
For example, Cov across all studied projects achieved by \utgen{} varies for different configurations: 35.74\% for \raggen{} with CL-7B, 29.98\% for \raggen{} with DS-7B, 32.20\% for \telpa{} with CL-7B, and 33.95\% for \telpa{} with DS-7B. In contrast, the average Cov achieved by Origin is 50.48\%, 50.07\%, 27.66\%, and 31.65\% for the same configurations, respectively.
The underperformance of \utgen{} is mainly due to its tendency to introduce incorrect semantic information, such as altering original test functionalities or introducing invalidity, which misleads the LLM during the unit test generation process.

\begin{tcolorbox}\textbf{RQ2 Summary:}
Incorporating \tech{}-refined test examples into prompts enhances ICL-based unit test generation, 
with an average improvement of 10.07\%/13.88\%/20.71\% and 25.97\%/28.22\%/45.99\% compared to using original and \utgen{}-refined test examples in terms of CSR/PR/Cov, respectively.

\end{tcolorbox}

\subsection{RQ3: Ablation Study}
\label{sec:ablation}

\subsubsection{Process}

In this RQ, we conducted an ablation study to examine the contribution of each core component (i.e., test purification and program-analysis-based post-processing) to the overall effectiveness of \tech{}.
Accordingly, we constructed three variants of \tech{}: 
\begin{itemize}
    \item \noslicing{}, which removes the component of test purification from \tech{}.
    That is, it directly performs textual clarity enhancement on the original unit tests rather than the purified ones.
    
    \item \nopa{}, which 
    removes the component of program-analysis-based post-processing from \tech{}.
    That is, it instructs the LLM to enhance textual clarity on each purified unit test and then directly utilizes the LLM-generated test as the refined one.
    
    \item \noboth{}, which removes both components (i.e., test purification and program-analysis-based post-processing).
    The prompting process for \noboth{} is similar to \utgen{}'s, but the former directly adopts the initial LLM-generated test as the refined version while the latter includes an iterative refinement process with LLMs.
\end{itemize}
We evaluated these variants in two scenarios: (1) refining unit tests (as in RQ1), and (2) using refined unit tests to enhance ICL-based unit test generation (as in RQ2).

\subsubsection{Results}

\begin{table*}[t]
\caption{Comparison between \tech{} and its variants in test-effectiveness-preserving degree (RQ3)}
\large
\centering
\renewcommand{\arraystretch}{1}
\centering
\label{tab:rq3_1_function}
\resizebox{0.65\linewidth}{!}{
\begin{threeparttable}
\begin{tabular}{l|rrrr|rrrr}
\toprule
\multirow{2}{*}{\textbf{Technique}} & \multicolumn{4}{c|}{\textbf{Developer-written Tests}} & \multicolumn{4}{c}{\textbf{Tool-generated Tests}} \\
                               & \multicolumn{1}{c}{\textbf{CSR}} & \multicolumn{1}{c}{\textbf{PR}} & \multicolumn{1}{c}{\textbf{Cov}} & \multicolumn{1}{c|}{\textbf{MS}} & \multicolumn{1}{c}{\textbf{CSR}} & \multicolumn{1}{c}{\textbf{PR}} & \multicolumn{1}{c}{\textbf{Cov}} &
                               \multicolumn{1}{c}{\textbf{MS}} \\
\midrule
Origin                & 100.00\%    & 99.75\%    & 94.07\%  & 73.97\%   & 100.00\%    & 100.00\%    & 43.43\%  & 57.21\%   \\
$\Delta$\tech{}                 & 0.00\%   & 0.00\%   & 0.00\% & 0.00\% & 0.00\%             & 0.00\%   & 0.00\%   & 0.00\%              \\
$\Delta$\noslicingname{}           & 0.00\%   & 0.00\% & 0.00\% & 0.00\%  & 0.00\%             & 0.00\%   & 0.00\%   & 0.00\%              \\
$\Delta$\nopaname{}               & -19.66\% & -21.29\% & -17.30\% & -2.34\%           & -5.26\%  & -6.01\%  & -1.24\%     & -0.76\%        \\
$\Delta$\nobothname{} & -23.10\% & -25.38\% & -17.91\%   & -4.18\%         & -12.22\% & -12.87\% & -5.49\%    & -0.86\%          \\
\bottomrule
\end{tabular}
\end{threeparttable}
}
\end{table*}

\begin{figure}[t]
\setlength{\abovecaptionskip}{-0.04cm} 
    \centering
        \includegraphics[width=0.92\linewidth]{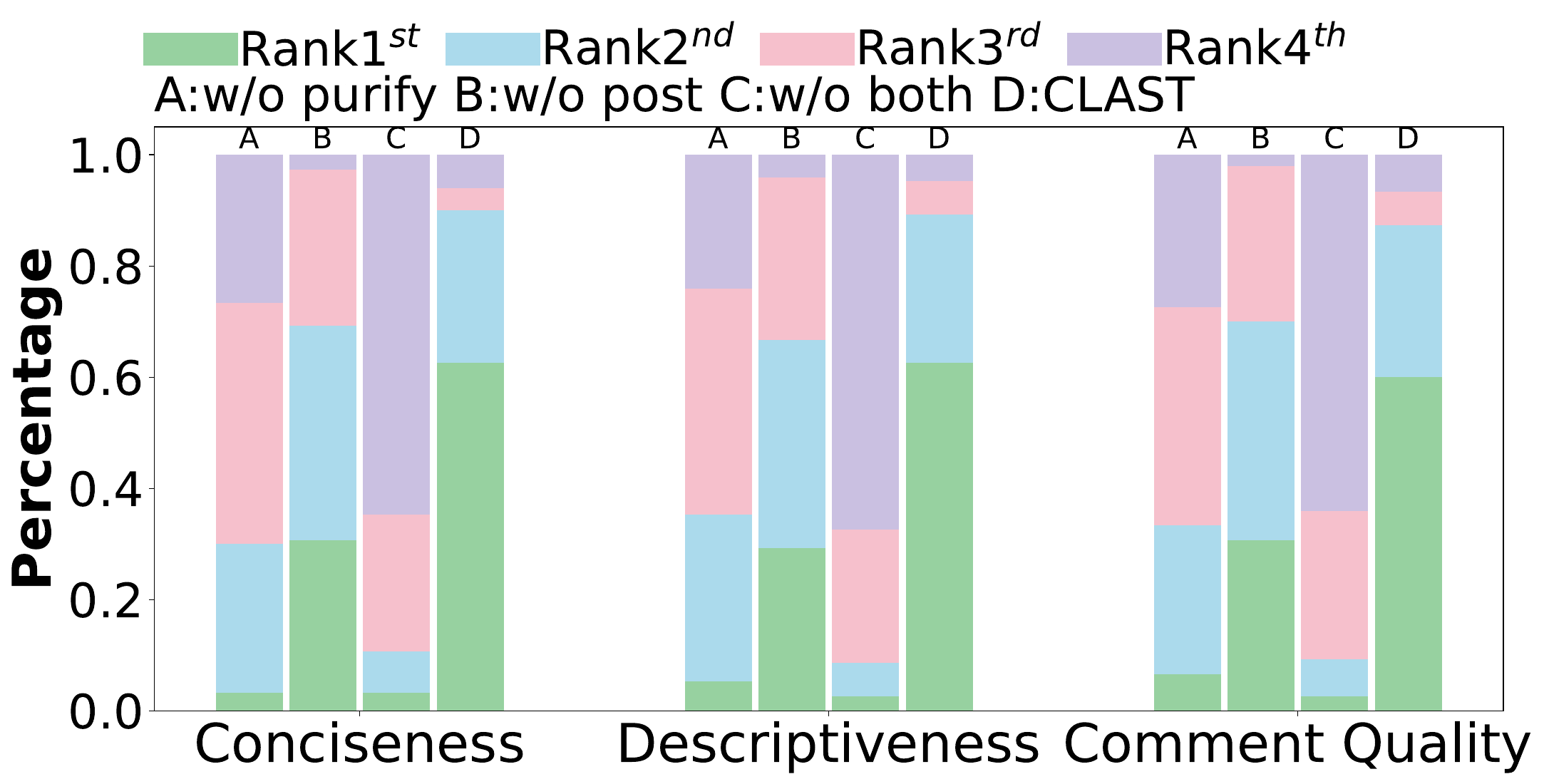}
        \caption{Comparison between \tech{} and its variants in terms of semantic clarity (RQ3)}
        \label{fig:rq3_user}
        \vspace{-1em}
\end{figure}

\begin{table*}[t]
\caption{Comparison between \tech{} and its variants in terms of the effectiveness for ICL-based test generation (RQ3)}
\label{tab:rq3_2_function}
\centering
\renewcommand{\arraystretch}{1.2}
\resizebox{0.78\linewidth}{!}{
\begin{threeparttable}
\begin{tabular}{l|l|rrr|rrr|rrr}
\toprule
\multicolumn{1}{c|}{\multirow{3}{*}{\textbf{LLM}}} & \multicolumn{1}{c|}{\multirow{3}{*}{\textbf{Technique}}} & \multicolumn{6}{c|}{\textbf{\raggen{}}}  & \multicolumn{3}{c}{\textbf{\telpa{}}} \\
\multicolumn{1}{c|}{}                                & \multicolumn{1}{c|}{}                               & \multicolumn{3}{c|}{\textbf{Open-source Projects}} & \multicolumn{3}{c|}{\textbf{Industrial Projects}} & \multicolumn{3}{c}{\textbf{Open-source Projects}} \\
\multicolumn{1}{c|}{}                                & \multicolumn{1}{c|}{}                               & \multicolumn{1}{c}{\textbf{CSR}} & \multicolumn{1}{c}{\textbf{PR}} & \multicolumn{1}{c|}{\textbf{Cov}} & \multicolumn{1}{c}{\textbf{CSR}} & \multicolumn{1}{c}{\textbf{PR}} & \multicolumn{1}{c|}{\textbf{Cov}} & \multicolumn{1}{c}{\textbf{CSR}} & \multicolumn{1}{c}{\textbf{PR}} & \multicolumn{1}{c}{\textbf{Cov}} \\
\midrule
\multirow{4}{*}{CL-7B}                              & \tech{}                                               & \textbf{67.05\%}                              & \textbf{36.60\%}                           & \textbf{65.37\%}                               & \textbf{61.46\%}               & \textbf{37.66\%}                & \textbf{45.91\%}                           & \textbf{86.00\%}                              & \textbf{62.01\%}                            & \textbf{37.52\%}                               \\
                                                    & \noslicingname{}                                    & 62.72\%                              & 35.00\%                           & 60.82\%                               & 58.54\%               & 36.26\%                & 42.39\%                           & 84.33\%                              & 48.82\%                            & 37.39\%                               \\
                                                    &  \nopaname{} & 64.02\%                              & 34.19\%                           & 57.20\%                               & 49.97\%               & 34.87\%                & 38.68\%                           & 84.23\%                              & 60.23\%                            & 36.28\%                               \\
                                                    & \nobothname{}                           & 58.97\%                              & 34.10\%                           & 55.91\%                               & 51.79\%               & 34.64\%                & 40.39\%                           & 76.70\%                              & 56.21\%                            & 29.25\%                               \\
\midrule
\multirow{4}{*}{DS-7B}                              & \tech{}                                               & \textbf{72.26\%}               
               & \textbf{47.37\%}                           & \textbf{70.28\%}                               & \textbf{59.49\%}            
    & \textbf{42.94\%}                & \textbf{47.98\%}                           & \textbf{85.82\%}                              
& \textbf{55.19\%}                            & \textbf{35.73\%}                               \\
                                                    & \noslicingname{}                                    & 68.90\%                              & 39.56\%                           & 65.79\%                    
           & 51.50\%               & 36.69\%                & 40.39\%                           & 84.83\%                              & 53.80\%                            & 35.50\%                               \\
                                                    & \nopaname{}                                             & 61.59\%                              & 38.66\%                           & 57.73\%                    
           & 52.16\%               & 35.49\%                & 41.86\%                           & 82.16\%                              & 52.95\%                            & 33.96\%                               \\
                                                    & \nobothname{}                           & 65.48\%                              & 36.35\%                           & 55.99\%                    
           & 48.64\%               & 34.32\%                & 39.11\%                           & 82.75\%                              & 52.95\%                            & 29.98\%                               \\
\bottomrule
\end{tabular}
\end{threeparttable}
}
\end{table*}

Table~\ref{tab:rq3_1_function} compares the test-effectiveness-preserving degree across \tech{} and its variants. The program-analysis-based post-processing component is crucial for preserving test effectiveness. The \nopa{} variant significantly degrades effectiveness, with average decreases of 12.46\%, 13.65\%, 9.27\%, and \ins{1.55\%} in CSR, PR, Cov, and \ins{MS} respectively, due to LLM hallucinations. The \noboth{} variant, which removes both purification and post-processing, further damages effectiveness, highlighting the importance of simplifying tests before refinement. Notably, \noslicing{} achieves the same results as \tech{} because it retains post-processing, ensuring that the effectiveness of the test cases remains intact.

Figure~\ref{fig:rq3_user} shows the semantic clarity results. Test purification significantly enhances clarity, as \noslicing{} and \noboth{} receive fewer first-place rankings and more last-place rankings compared to \tech{} and \nopa{}. This demonstrates that purification improves LLMs' comprehension, leading to more accurate comments and identifiers.

Table~\ref{tab:rq3_2_function} evaluates the impact of refined tests on ICL-based test generation. \tech{} consistently outperforms its variants across all metrics, ICL-based unit test generation approaches (\raggen{} and \telpa{}), LLMs (CL-7B and DS-7B), and project types (open-source and industrial projects). Removing post-processing reduces CSR, PR, and Cov by \ins{8.78}\%, \ins{9.01}\%, and \ins{12.25}\%, on average across all ICL-based approaches, LLMs, and projects, while removing purification reduces them by \ins{4.92}\%, \ins{11.23}\%, and \ins{6.77}\%. 
The results confirm that both test effectiveness preservation (contributed by program-analysis-based post-processing) and clarity enhancement (contributed by test purification) are crucial to the capability of \tech{} in improving ICL-based test generation.

\begin{tcolorbox}\textbf{RQ3 Summary:}
Both core components (i.e., test purification and program-analysis-based post-processing) contribute significantly to \tech{}'s effectiveness, which improves both the quality of refined tests and the effectiveness of ICL-based unit test generation.
\end{tcolorbox}

\section{Discussion}
\label{sec:dis}

\textbf{Perceptions on \tech{}.}
 To further assess the practical value of \tech{}, we asked user study participants to rank their willingness to incorporate refined tests into real-world projects. \tech{}-refined tests received the highest first-place ranking (90.67\%), compared to \utgen{} (8.00\%) and original tests (1.33\%), confirming their practicality and effectiveness.
Participants also provided insights on desirable test characteristics. We analyzed their answers using a card sorting method~\cite{cardsort}. The two key insights are summarized:
\begin{itemize}[leftmargin=15pt]
    \item \textit{Concise Scenarios and Clear Textual Clarity}: 80.0\% of participants emphasized that concise scenarios and straightforward naming conventions improve comprehension for both developers and LLMs. \tech{}-refined tests were praised for their clarity and structure, which reduce ambiguity and enhance LLM-friendliness.
 
    \item \textit{Avoiding Excessive Documentation}: While necessary comments are valuable, 33.3\% of participants mentioned overly detailed documentation, which can complicate tests and affect semantic clarity. 
    This necessitates adopting a more adaptive approach such as incorporating a complexity-based filtering mechanism to avoid redundancy.
    
\end{itemize}

\smallskip
\noindent
\textbf{Efficiency of \tech{}.}
While \tech{} demonstrates remarkable performance, it is essential to consider the trade-off between its effectiveness and cost. 
Specifically, we measured the time efficiency of both \tech{} and \utgen{}. 
The results indicate that \utgen{} takes an average of 93.327s to refine a single unit test, whereas \tech{} requires only 55.133s (i.e., allocating 0.003s for test purification and 55.13s for textual clarity enhancement). 
The reason may be that \utgen{} iteratively invokes the LLM for refinement, but \tech{} only requires a single round of invocation.
Moreover, the potential for parallel execution could further accelerate the refinement process of \tech{}, boosting its practicality.
These findings suggest that \tech{} offers a significantly more efficient and scalable refinement process.

\begin{table}[t]
\caption{Impact of \tech{} on HITS Performance}
\centering
\label{tab:hits_res}
\resizebox{0.9\linewidth}{!}{
\begin{threeparttable}
\begin{tabular}{l|rrrr}
\toprule
\textbf{Metric} & \multicolumn{1}{l}{\textbf{w/o RAG}} & \multicolumn{1}{l}{\textbf{Origin}} & \multicolumn{1}{l}{\textbf{\utgen{}}} & \multicolumn{1}{l}{\textbf{\tech{}}} \\
\midrule
CSR    & 32.37\%                                  & 32.46\%                             & 27.31\%                            & \textbf{34.27\%}                                  \\
PR     & 9.99\%                                   & 14.59\%                             & 14.13\%                            & \textbf{16.18\%}                                  \\
Cov    & 21.60\%                                  & 39.19\%                             & 38.50\%                            & \textbf{41.90\%}                       \\
\bottomrule
\end{tabular}
\end{threeparttable}
}
\end{table}

\begin{table}[t]
\caption{Generalizability of \tech{} on larger LLMs}
\label{tab: larger_llm}
\centering
\renewcommand{\arraystretch}{1.2}
\resizebox{0.95\linewidth}{!}{
\begin{threeparttable}
\begin{tabular}{l|rrr|rrr}
\toprule
 \multirow{2}{*}{\textbf{Technique}} & \multicolumn{3}{c|}{\textbf{Open source projects}}                                                     & \multicolumn{3}{c}{\textbf{Industrial projects}}                                                      \\
                                    & \multicolumn{1}{c}{\textbf{CSR}} & \multicolumn{1}{c}{\textbf{PR}} & \multicolumn{1}{c|}{\textbf{Cov}} & \multicolumn{1}{c}{\textbf{CSR}} & \multicolumn{1}{c}{\textbf{PR}} & \multicolumn{1}{c}{\textbf{Cov}} \\
\hline
\textbf{Base}       & 68.34\%              & 61.99\%          & 75.96\%          & 70.25\%             & 71.42\%          & 47.54\%          \\
                                \textbf{Origin}     & 75.76\%              & 67.94\%          & 87.66\%          & 75.31\%             & 72.84\%          & 48.14\%          \\
                                \textbf{UTgen}      & 70.62\%              & 63.77\%          & 77.61\%          & 70.12\%             & 65.28\%          & 45.08\%          \\
                                \textbf{CLAST}      & \textbf{79.31\%}     & \textbf{71.44\%} & \textbf{90.09\%} & \textbf{78.21\%}    & \textbf{73.98\%} & \textbf{50.12\%} \\
                                      \bottomrule
\end{tabular}
\end{threeparttable}
}
\end{table}

\smallskip
\noindent
\textbf{Generalizability.}
To highlight the benefit of ICL in LLM-based test generation and \tech{}'s generalizability, we integrated \tech{} into HITS~\cite{wang2024hits}, a state-of-the-art LLM-based test generation approach that originally lacks in-context examples. 
We enhanced HITS with Retrieval-Augmented Generation (RAG) to retrieve similar methods and unit tests for in-context learning.
\ins{
Specifically, we adopted RAGGen’s well-established RAG approach.
}
We evaluated HITS in four settings: without RAG, with original test examples, with \utgen{}-refined examples, and with \tech{}-refined examples, using four open-source projects from Defects4J and DS-7B as the representative. 
Table~\ref{tab:hits_res} shows the average results across all projects. From the table, incorporating in-context examples indeed improves HITS' effectiveness, underscoring the importance of in-context examples. 
Notably, \tech{}-refined examples consistently outperform both the original and \utgen{}-refined examples, demonstrating \tech{}'s effectiveness and generalizability in enhancing LLM-based unit test generation.

\ins{
To investigate \tech{}'s generalizability across LLM scales, we evaluated DeepSeek-V3 (610B), a state-of-the-art model surpassing GPT-4 on code/math tasks~\cite{liu2024deepseek} on RAGGen. The results are summarized in Table~\ref{tab: larger_llm}. From the results, our core findings are reinforced: while larger models perform better, \tech{}-refined examples consistently enhance performance, improving Base/Origin/UTgen by 13.69\%/4.27\%/11.92\% in CSR, 9.41\%/3.26\%/12.68\% in PR, and 12.01\%/3.44\%/13.63\% in Cov. Though improvements are smaller than with 7B models, likely due to the already high performance of larger models (e.g., DeepSeek-V3 alone improves Cov by 28.19\% over DS-7B), \tech{} still boosts performance, demonstrating its generalizability across LLM scales. 
}

\ins{
Regarding the generalizability to different languages and real-world settings. \tech{}'s idea is language-agnostic. Adapting it involves replacing language-specific implementations (e.g., AST parsing) and refining prompts to match the target language's syntax and testing conventions. Tools like Tree-sitter (for JavaScript/Python/C++) provide support for program analysis, while prompt adjustments just require updating examples to align with frameworks like unittest. This makes \tech{}'s extension highly feasible. Additionally, \tech{} takes an average of 55.13 seconds per test, with most of the overhead coming from LLM inference rather than static analysis/transformation. This is comparable to existing LLM-based tools like Copilot, which are already integrated into real-world workflows as asynchronous aids, not in latency-critical paths. This makes \tech{} similarly deployable.
}

\smallskip
\noindent
\textbf{Future Work.}
Several promising directions include: 
(I) Mitigating the risk of excessive documentation by guiding LLMs to insert comments at key points, identified based on factors such as statement complexity and importance.
(II) 
Incorporating runtime information, such as test coverage, to enhance naming and commenting by clarifying each test’s purpose. 
(III) Extending \tech{} to enhance supervised fine-tuning of LLM-based test generation by refining training data
and provide high-quality unit tests to aid developers in software debugging and maintenance.

\section{Threats to Validity}
The threats to \textbf{external} validity mainly lie in subject selection and studied ICL-based unit test generation approaches. To mitigate these, we evaluated \tech{} on both open-source and industrial projects, and integrated it with two advanced ICL-based approaches, \raggen{} and \telpa{}, covering diverse test example usage scenarios. 
In future work, we can extend \tech{} to more approaches, including those without test example incorporation originally.  

The threat to \textbf{internal} validity primarily lies in the implementation. \tech{} underwent rigorous code review and testing by three authors. For \utgen{}, \raggen{}, and \telpa{}, we used their publicly released artifacts~\cite{utgen_page, rag_page, telpa_page}. 
LLMs (CL-7B, DS-7B from Hugging Face~\cite{huggingface} and DeepSeek-V2.5 API~\cite{deepseek}) were used following their official guidelines.

The threats to \textbf{construct} validity include LLM randomness, data leakage, metric selection, and potential semantic drift. Following the existing work, we set LLM temperature to zero and repeated all quantitative experiments for 10 times and reported the average results to reduce randomness~\cite{telpa, rted, ase_empirical, chen2017learning, d3}. To prevent data leakage, we used internal industrial projects and tool-generated tests free from LLM training data. To address the metric threat, we employed diverse metrics (CSR, PR, Cov, \ins{and MS}) to comprehensively evaluate \tech{}'s effectiveness.
\ins{
To address the final threat, \tech{} uses carefully-designed prompts with task-specific examples to generate identifiers and comments that preserve code semantics. Though minor semantic drift (e.g., mColumn3→expectedColumnMatrix) may occur, this is mitigated via complementary reinforcement: descriptive comments clarify identifiers (e.g., ``expected column matrix for column 3''), and vice versa. Additionally, embedding-based similarity checks further align comments with intent.
}

\section{Related Work}
\ins{
\textbf{Test Purification.}
Xuan and Martin~\cite{purification} first introduced test purification, which reduces the size of a test case via program slicing from the assertion to improve fault localization. Their approach extracts minimal failure-inducing statement subsets using dynamic slicing.
In contrast, CLAST enhances semantic clarity for LLMs by producing each example that focuses on a clear scenario with thorough intra-scenario coverage, making the example focused yet comprehensive for LLMs to learn high-quality testing patterns. Since LLMs learn statically, aggressive pruning (as in dynamic slicing) may remove useful structural or semantic cues. Static slicing is also lighter and avoids runtime instrumentation.
}

\textbf{Test Quality Improvement.}
Prior work has focused on improving test quality, primarily addressing test smells and readability. Studies like Soares et al.~\cite{tse_smell} and Peruma et al.~\cite{peruma2019distribution} explored strategies to reduce test smells, while Lucas et al.~\cite{lucas2024evaluating} and Gao et al.~\cite{UTRefactor} investigated LLMs for detecting and repairing test smells. Other efforts, such as Zhang et al.~\cite{zhang2016towards} and Daka et al.~\cite{Daka}, aimed to enhance readability by generating descriptive test names or summaries. 
Recently, Deljouyi et al.~\cite{utgen} introduced \utgen{}, leveraging LLMs to improve semantic clarity. However, it struggles with complex test scenarios and LLM hallucinations. Unlike them, \tech{} proposes a novel two-step approach that purifies complex tests by splitting them into clearer ones and refines their semantic clarity by leveraging LLMs and program analysis.

\smallskip
\noindent
\textbf{LLM-based Unit Test Generation.}
LLM-based test generation approaches fall into two categories: training-based and prompting-based. Training-based methods, such as ATHENATEST~\cite{athenatest} and A3Test~\cite{a3test}, train LLMs on large datasets of unit tests, achieving strong results but requiring significant resources. Prompting-based approaches, like ChatTester~\cite{chattester}, SymPrompt~\cite{symprompt}, and HITS~\cite{wang2024hits}, use contextual prompts to guide LLMs in generating tests, offering flexibility and reduced reliance on fine-tuning. 
Recent advancements, including \raggen{} and \telpa{}, demonstrate the effectiveness of incorporating in-context-learning (i.e., providing in-context test examples in prompts) for prompting-based test generation. 
\textit{
Unlike these approaches, which focus on context construction or example selection, 
\tech{} enhances in-context-learning effectiveness by improving the semantic clarity of test examples.
}
In general, our work is orthogonal to existing methods and can be integrated to further improve their performance.

\section{Conclusion}
Existing ICL-based unit test generation techniques are hindered by the limited semantic clarity of unit test examples.
To tackle this issue, we developed \tech{}, an innovative refinement technique that enhances the quality of unit test examples by splitting a complex test into a set of purified ones and improving their textual clarity using a combination of LLMs and program analysis.
Our extensive evaluation on real-world projects demonstrates that \tech{} significantly outperforms the state-of-the-art test refinement technique \utgen{} in both preserving test effectiveness and enhancing the semantic clarity of unit tests. Results also show that incorporating \tech{}-refined unit tests can effectively enhance LLM-based unit test generation, i.e., \raggen{} and \telpa{}.

\section*{Acknowledgment}
We thank all the ASE anonymous reviewers for their valuable comments.
This work was supported by the National Key Research and Development Program of China (Grant No. 2024YFB4506300), the National Natural Science Foundation of China (Grant Nos. 62322208, 62232001), and the Emerging Frontiers Cultivation Program of Tianjin University Interdisciplinary Center.

\balance
\bibliographystyle{IEEEtran}
\bibliography{reference}

\begin{thebibliography}{10}
\providecommand{\url}[1]{#1}
\csname url@samestyle\endcsname
\providecommand{\newblock}{\relax}
\providecommand{\bibinfo}[2]{#2}
\providecommand{\BIBentrySTDinterwordspacing}{\spaceskip=0pt\relax}
\providecommand{\BIBentryALTinterwordstretchfactor}{4}
\providecommand{\BIBentryALTinterwordspacing}{\spaceskip=\fontdimen2\font plus
\BIBentryALTinterwordstretchfactor\fontdimen3\font minus \fontdimen4\font\relax}
\providecommand{\BIBforeignlanguage}[2]{{%
\expandafter\ifx\csname l@#1\endcsname\relax
\typeout{** WARNING: IEEEtran.bst: No hyphenation pattern has been}%
\typeout{** loaded for the language `#1'. Using the pattern for}%
\typeout{** the default language instead.}%
\else
\language=\csname l@#1\endcsname
\fi
#2}}
\providecommand{\BIBdecl}{\relax}
\BIBdecl

\bibitem{pynguin}
S.~Lukasczyk and G.~Fraser, ``Pynguin: Automated unit test generation for python,'' in \emph{Proceedings of the ACM/IEEE 44th International Conference on Software Engineering: Companion Proceedings}, 2022, pp. 168--172.

\bibitem{evosuite}
G.~Fraser and A.~Arcuri, ``Evosuite: automatic test suite generation for object-oriented software,'' in \emph{Proceedings of the 19th ACM SIGSOFT symposium and the 13th European conference on Foundations of software engineering}, 2011, pp. 416--419.

\bibitem{randoop}
C.~Pacheco and M.~D. Ernst, ``Randoop: feedback-directed random testing for java,'' in \emph{Companion to the 22nd ACM SIGPLAN conference on Object-oriented programming systems and applications companion}, 2007, pp. 815--816.

\bibitem{ase_empirical}
L.~Yang, C.~Yang, S.~Gao, W.~Wang, B.~Wang, Q.~Zhu, X.~Chu, J.~Zhou, G.~Liang, Q.~Wang, and J.~Chen, ``On the evaluation of large language models in unit test generation,'' in \emph{Proceedings of the 39th IEEE/ACM International Conference on Automated Software Engineering}, 2024, p. 1607–1619.

\bibitem{telpa}
C.~Yang, J.~Chen, B.~Lin, Z.~Wang, and J.~Zhou, ``Advancing code coverage: Incorporating program analysis with large language models,'' \emph{ACM Transactions on Software Engineering and Methodology}, 2024.

\bibitem{utgen}
A.~Deljouyi, R.~Koohestani, M.~Izadi, and A.~Zaidman, ``Leveraging large language models for enhancing the understandability of generated unit tests,'' in \emph{{ICSE}}.\hskip 1em plus 0.5em minus 0.4em\relax {ACM}, 2024.

\bibitem{dan}
C.~Yang, J.~Chen, J.~Jiang, and Y.~Huang, ``Dependency-aware code naturalness,'' \emph{Proceedings of the ACM on Programming Languages}, vol.~8, no. OOPSLA2, pp. 2355--2377, 2024.

\bibitem{homepage}
``Project homepage,'' 2025, \url{https://github.com/chenyangyc/CLAST}.

\bibitem{ahmed2021slicer4j}
K.~Ahmed, M.~Lis, and J.~Rubin, ``Slicer4j: a dynamic slicer for java,'' in \emph{Proceedings of the 29th ACM Joint Meeting on European Software Engineering Conference and Symposium on the Foundations of Software Engineering}, 2021, pp. 1570--1574.

\bibitem{chattester}
Z.~Yuan, Y.~Lou, M.~Liu, S.~Ding, K.~Wang, Y.~Chen, and X.~Peng, ``No more manual tests? evaluating and improving chatgpt for unit test generation,'' \emph{arXiv preprint arXiv:2305.04207}, 2023.

\bibitem{testdescriber}
S.~Panichella, A.~Panichella, M.~Beller, A.~Zaidman, and H.~C. Gall, ``The impact of test case summaries on bug fixing performance: An empirical investigation,'' in \emph{Proceedings of the 38th international conference on software engineering}, 2016, pp. 547--558.

\bibitem{Daka}
E.~Daka, J.~M. Rojas, and G.~Fraser, ``Generating unit tests with descriptive names or: would you name your children thing1 and thing2?'' in \emph{{ISSTA}}.\hskip 1em plus 0.5em minus 0.4em\relax {ACM}, 2017, pp. 57--67.

\bibitem{deeptc}
D.~Roy, Z.~Zhang, M.~Ma, V.~Arnaoudova, A.~Panichella, S.~Panichella, D.~Gonzalez, and M.~Mirakhorli, ``Deeptc-enhancer: Improving the readability of automatically generated tests,'' in \emph{Proceedings of the 35th IEEE/ACM International Conference on Automated Software Engineering}, 2020, pp. 287--298.

\bibitem{wei2023automatically}
C.~Wei, L.~Xiao, T.~Yu, X.~Chen, X.~Wang, S.~Wong, and A.~Clune, ``Automatically tagging the “aaa” pattern in unit test cases using machine learning models,'' \emph{IEEE Transactions on Software Engineering}, vol.~49, no.~5, pp. 3305--3324, 2023.

\bibitem{gao2019teccd}
Y.~Gao, Z.~Wang, S.~Liu, L.~Yang, W.~Sang, and Y.~Cai, ``Teccd: A tree embedding approach for code clone detection,'' in \emph{2019 IEEE international conference on software maintenance and evolution (ICSME)}.\hskip 1em plus 0.5em minus 0.4em\relax ieee, 2019, pp. 145--156.

\bibitem{oh2024csa}
S.~Oh and S.~Yoo, ``Csa-trans: Code structure aware transformer for ast,'' \emph{arXiv preprint arXiv:2404.05767}, 2024.

\bibitem{codebleu}
S.~Ren, D.~Guo, S.~Lu, L.~Zhou, S.~Liu, D.~Tang, N.~Sundaresan, M.~Zhou, A.~Blanco, and S.~Ma, ``Codebleu: a method for automatic evaluation of code synthesis,'' \emph{arXiv preprint arXiv:2009.10297}, 2020.

\bibitem{chatunitest}
Z.~Xie, Y.~Chen, C.~Zhi, S.~Deng, and J.~Yin, ``Chatunitest: a chatgpt-based automated unit test generation tool,'' \emph{arXiv preprint arXiv:2305.04764}, 2023.

\bibitem{junit}
``Junit,'' 2024, \url{https://junit.org/junit5/}.

\bibitem{coderujb}
Z.~Zeng, Y.~Wang, R.~Xie, W.~Ye, and S.~Zhang, ``Coderujb: An executable and unified java benchmark for practical programming scenarios,'' in \emph{Proceedings of the 33rd ACM SIGSOFT International Symposium on Software Testing and Analysis}, 2024, pp. 124--136.

\bibitem{defects4j}
R.~Just, D.~Jalali, and M.~D. Ernst, ``Defects4j: A database of existing faults to enable controlled testing studies for java programs,'' in \emph{Proceedings of the 2014 international symposium on software testing and analysis}, 2014, pp. 437--440.

\bibitem{symprompt}
G.~Ryan, S.~Jain, M.~Shang, S.~Wang, X.~Ma, M.~K. Ramanathan, and B.~Ray, ``Code-aware prompting: A study of coverage-guided test generation in regression setting using llm,'' \emph{Proceedings of the ACM on Software Engineering}, vol.~1, no. FSE, pp. 951--971, 2024.

\bibitem{chen2023toward}
J.~Chen, Y.~Liang, Q.~Shen, J.~Jiang, and S.~Li, ``Toward understanding deep learning framework bugs,'' \emph{ACM Transactions on Software Engineering and Methodology}, vol.~32, no.~6, pp. 1--31, 2023.

\bibitem{huggingface_leaderboard}
``Hugging face big code model leaderboard,'' 2024, \url{https://huggingface.co/spaces/bigcode/bigcode-models-leaderboard}.

\bibitem{codellama}
B.~Roziere, J.~Gehring, F.~Gloeckle, S.~Sootla, I.~Gat, X.~E. Tan, Y.~Adi, J.~Liu, T.~Remez, J.~Rapin \emph{et~al.}, ``Code llama: Open foundation models for code,'' \emph{arXiv preprint arXiv:2308.12950}, 2023.

\bibitem{deepseek_paper}
D.~Guo, Q.~Zhu, D.~Yang, Z.~Xie, K.~Dong, W.~Zhang, G.~Chen, X.~Bi, Y.~Wu, Y.~Li \emph{et~al.}, ``Deepseek-coder: When the large language model meets programming--the rise of code intelligence,'' \emph{arXiv preprint arXiv:2401.14196}, 2024.

\bibitem{tree-sitter}
``tree-sitter,'' 2024, \url{https://tree-sitter.github.io/tree-sitter}.

\bibitem{deepseek}
``Deepseek,'' 2024, \url{https://github.com/deepseek-ai/DeepSeek-Coder}.

\bibitem{cobertura}
``cobertura,'' 2024, \url{https://github.com/cobertura/cobertura}.

\bibitem{torch}
``Pytorch,'' 2024, \url{http://pytorch.org}.

\bibitem{transformers}
``Transformers,'' 2024, \url{https://github.com/huggingface/transformers}.

\bibitem{vllm}
W.~Kwon, Z.~Li, S.~Zhuang, Y.~Sheng, L.~Zheng, C.~H. Yu, J.~Gonzalez, H.~Zhang, and I.~Stoica, ``Efficient memory management for large language model serving with pagedattention,'' in \emph{Proceedings of the 29th Symposium on Operating Systems Principles}, 2023, pp. 611--626.

\bibitem{cardsort}
D.~Spencer and T.~Warfel, ``Card sorting: a definitive guide,'' \emph{Boxes and arrows}, vol.~2, no. 2004, pp. 1--23, 2004.

\bibitem{wang2024hits}
Z.~Wang, K.~Liu, G.~Li, and Z.~Jin, ``Hits: High-coverage llm-based unit test generation via method slicing,'' \emph{arXiv preprint arXiv:2408.11324}, 2024.

\bibitem{liu2024deepseek}
A.~Liu, B.~Feng, B.~Xue, B.~Wang, B.~Wu, C.~Lu, C.~Zhao, C.~Deng, C.~Zhang, C.~Ruan \emph{et~al.}, ``Deepseek-v3 technical report,'' \emph{arXiv preprint arXiv:2412.19437}, 2024.

\bibitem{utgen_page}
``Utgen repository,'' 2024, \url{https://github.com/amirdeljouyi/UTGen}.

\bibitem{rag_page}
``Raggen repository,'' 2024, \url{https://github.com/LeonYang95/LLM4UT}.

\bibitem{telpa_page}
``Telpa repository,'' 2025, \url{https://github.com/chenyangyc/TELPA}.

\bibitem{huggingface}
``Hugging face,'' 2024, \url{https://huggingface.co}.

\bibitem{rted}
C.~Yang, Z.~Wang, Y.~Jiang, L.~Yang, Y.~Zheng, J.~Zhou, and J.~Chen, ``Reflective unit test generation for precise type error detection with large language models,'' in \emph{Proceedings of the 40th IEEE/ACM International Conference on Automated Software Engineering}, 2025.

\bibitem{chen2017learning}
J.~Chen, Y.~Bai, D.~Hao, Y.~Xiong, H.~Zhang, and B.~Xie, ``Learning to prioritize test programs for compiler testing,'' in \emph{2017 IEEE/ACM 39th International Conference on Software Engineering (ICSE)}.\hskip 1em plus 0.5em minus 0.4em\relax IEEE, 2017, pp. 700--711.

\bibitem{d3}
C.~Yang, J.~Chen, X.~Fan, J.~Jiang, and J.~Sun, ``Silent compiler bug de-duplication via three-dimensional analysis,'' in \emph{Proceedings of the 32nd ACM SIGSOFT International Symposium on Software Testing and Analysis}, 2023, pp. 677--689.

\bibitem{purification}
J.~Xuan and M.~Monperrus, ``Test case purification for improving fault localization,'' in \emph{Proceedings of the 22nd ACM SIGSOFT international symposium on foundations of software engineering}, 2014, pp. 52--63.

\bibitem{tse_smell}
E.~Soares, M.~Ribeiro, R.~Gheyi, G.~Amaral, and A.~Santos, ``Refactoring test smells with junit 5: Why should developers keep up-to-date?'' \emph{IEEE Transactions on Software Engineering}, vol.~49, no.~3, pp. 1152--1170, 2022.

\bibitem{peruma2019distribution}
A.~Peruma, K.~S. Almalki, C.~D. Newman, M.~W. Mkaouer, A.~Ouni, and F.~Palomba, ``On the distribution of test smells in open source android applications: An exploratory study,'' 2019.

\bibitem{lucas2024evaluating}
K.~Lucas, R.~Gheyi, E.~Soares, M.~Ribeiro, and I.~Machado, ``Evaluating large language models in detecting test smells,'' \emph{arXiv preprint arXiv:2407.19261}, 2024.

\bibitem{UTRefactor}
Y.~Gao, X.~Hu, X.~Yang, and X.~Xia, ``Context-enhanced llm-based framework for automatic test refactoring,'' \emph{CoRR}, vol. abs/2409.16739, 2024.

\bibitem{zhang2016towards}
B.~Zhang, E.~Hill, and J.~Clause, ``Towards automatically generating descriptive names for unit tests,'' in \emph{Proceedings of the 31st IEEE/ACM International Conference on Automated Software Engineering}, 2016, pp. 625--636.

\bibitem{athenatest}
M.~Tufano, D.~Drain, A.~Svyatkovskiy, S.~K. Deng, and N.~Sundaresan, ``Unit test case generation with transformers and focal context,'' \emph{arXiv preprint arXiv:2009.05617}, 2020.

\bibitem{a3test}
S.~Alagarsamy, C.~Tantithamthavorn, and A.~Aleti, ``A3test: Assertion-augmented automated test case generation,'' \emph{arXiv preprint arXiv:2302.10352}, 2023.

\end{thebibliography}

\end{document}